# On-chip calibration of Microscale-Thermocouples for Precise Temperature Measurement

Hassan Irshad Bhatti

Advanced Semiconductor Laboratory, Electrical and Computer Engineering Program, CEMSE Division, King Abdullah University of Science and Technology (KAUST), Thuwal 23955-6900, Kingdom of Saudi Arabia hassan.bhatti@kaust.edu.sa,

## Abstract

Precise temperature measurement at micro/nanoscale is crucial across various domains including physical sciences, chemical processes, industrial production, medical diagnosis, weather forecasting, electronics, and biology. Micro/nanoscale thermal mapping requires precise techniques such as thermocouples, resistance-based devices, infrared thermography, optical interferometry, Raman thermometry, and Time domain-thermoreflectance (TDTR) method. Each method has its advantages and limitations, emphasizing the importance of selecting the appropriate technique. Among these methods, micro-thin film thermocouples (TFTCs) offer a compelling solution due to their direct contact-based temperature measurements, minimal surface preparation requirements, lower cost, and robustness against environmental factors. Thermocouples work on the well-established Seebeck effect, where a voltage is generated proportional to the temperature difference between two points. However, at micro/nanoscale, the Seebeck coefficients of thermocouples differ from those in bulk materials, requiring experimental calibration for precise measurements. To address this, we introduce an on-chip characterization platform with a differential temperature measurement setup on a borosilicate glass substrate. This platform utilizes a microheater as a localized heat source to elevate the temperature at the hot junction of the TFTC while maintaining the cold junction at ambient conditions. Numerical simulations are employed to engineer both the microheater and TFTC junction for precise temperature control. The functionality of this platform is validated by fabricating TFTCs using standard fabrication processes and measuring the TFTC response to determine the differential Seebeck coefficient of a Platinum-Chromium TFTC Junction. The calculated sensitivity of Pt/Cr TFTCs using this calibration method is 19.23 ± 0.405 µV/C.

## Introduction

Accurate temperature measurement is a fundamental requirement across various domains, spanning physical sciences, chemical processes [1], and beyond. Its significance extends to practical applications in industrial production, where precise temperature control ensures optimal manufacturing conditions [2, 3]. In medical diagnosis, temperature monitoring aids in identifying health anomalies and guiding treatment decisions. Additionally, weather forecasting relies on accurate temperature data to predict climate patterns. Furthermore, in the realms of electronics and biology, maintaining specific temperature ranges is critical for device performance and biological processes. Whether it's a semiconductor chip or an enzymatic reaction, temperature plays a pivotal role in scientific processes and phenomena [4].

For micro/nanoscale temperature mapping and thermal characterization, various measurement techniques are employed such as thermocouples [5, 6] infrared thermography [7, 8], optical interferometry, Raman thermometry [9, 10], and Time domain-thermoreflectance method. While optical approaches allow non-invasive temperature sensing, the diffraction of light fundamentally limits their resolutions, and their use can be challenging in a light-scattering environment. Infrared thermography offers non-contact temperature measurements with quick detection over large areas but struggles with reflective surfaces and emissivity variations and spatial resolution [11, 12]. Optical Interferometry provides high precision and sensitivity but is limited by the need for vibration-free environments and transparent materials [13, 14]. Raman Thermometry excels in sensitivity, yet it demands complex, expensive equipment and faces challenges with sample fluorescence [11]. Thermoreflectance enables high spatial resolution on surfaces but requires complex calibration, is affected by material properties, and light sources, and is highly expensive. Each technique, while advantageous for specific applications, suffers from limitations regarding accuracy, physical nature of materials, applicability, and environmental sensitivity, highlighting the importance of selecting the appropriate method for micro/nanoscale temperature measurement. Micro-thin film thermocouples (TFTCs) present a compelling solution amidst the limitations of other temperature measurement techniques, offering a unique blend of advantages. Unlike methods such as infrared thermography and Raman thermometry, which struggle with surface emissivity variations and require expensive setups, TFTCs provide direct, contact-based temperature measurements with minimal surface preparation and at a lower cost [11, 15]. They bypass the need for optical access and are not affected by ambient light conditions, addressing key limitations of thermoreflectance. Unlike optical interferometry, TFTCs are robust against environmental vibrations and do not require transparent materials, making them versatile across a wide range of applications. Their ability to integrate directly onto surfaces for localized temperature sensing, coupled with their high sensitivity and response times, positions TFTCs as an excellent solution for precise temperature measurement in applications including but not limited to electronics, energy systems, and material sciences.

Thermocouples are temperature-sensing devices that consist of two wires, each made of distinct metals/metal alloys, joined to form a junction. The electrical properties of these alloys change with temperature, resulting in a voltage proportional to the temperature difference between the wires. By measuring this voltage, the junction temperature can be accurately determined. Thermocouples are passive devices which means that they do not need external power to function. Thermocouples find widespread use in temperature sensing due to their straightforward design, durability, and capability to measure a broad range of temperatures[16-19]. The working principle of thermocouples is based on the Seebeck effect. The Seebeck effect, a fundamental phenomenon in thermoelectricity, arises when an electrically conducting material exhibits a voltage difference between two points at different temperatures. Specifically, when a temperature gradient exists across the material, an electromotive force (EMF) emerges, resulting in a measurable voltage. Mathematically, the relationship between the change in voltage (∇V) and the temperature change (∇T) is expressed as:

$$\nabla V = -S(T)\nabla T$$

where *S(T)* represents the Seebeck coefficient, a material-specific property that characterizes the thermoelectric behavior. This coefficient quantifies the voltage generated per unit temperature difference and plays a crucial role in designing thermocouples and thermoelectric devices [20].

One limitation of using thermocouples at micro/nanoscale is that their Seebeck coefficients, $S_v$, differ from those in bulk materials when scaled down to such dimensions [21]. Consequently, calibrating these micro TFTCs and refining their manufacturing process necessitates an experimental approach to precisely measure their differential Seebeck coefficients. This measurement demands an accurate assessment of the temperature difference across the thermocouple's hot and cold junctions and the resulting open-circuit voltage. In traditional thermocouples, this temperature gradient is typically established by immersing the cold junction in an ice bath and heating the hot junction using a hot plate, muffle furnace or a hot tube furnace [22]. However, the very small size of micro-TFTC means that using a hot or cold bath would result in a uniform temperature change throughout the device especially when substrates such Silicon, [22-24] are used, making this method ineffective. Furthermore, some studies have used resistance based sensors for baseline calibration in a hot furnace and then subsequently use it to calibrate micro-TFTC. In most cases, external cooling has been utilized to keep the cold ends at a fixed temperature which comes at an extra cost and auxiliary equipment. Therefore, to calibrate the micro-TFTC a novel technique is required which does not cause the uniform temperature change and enables on-chip calibration of such devices while also not requiring external heating or cooling.

Hence, we introduce an on-chip characterization platform equipped with a differential temperature measurement setup on a borosilicate glass substrate. This setup uses a microheater as a localized heat source to selectively elevate the temperature at the hot junction, while the cold junction is maintained at ambient conditions due to the low thermal conductivity of the substrate. Our design does not require the need of external cooling or heating to achieve the hot and cold zones. To achieve precise temperature measurement at the thermocouple's hot junction, both the microheater and TFTC junction are engineered, utilizing numerical simulations to achieve a hot junction at elevated temperature and a cold zone at ambient temperature. To validate the functionality of this characterization platform, we fabricate the device using standard fabrication processes and measure the TFTC response to determine the differential Seebeck coefficient of a Platinum-Chromium (Pt/Cr) TFTC Junction.

## Experimental Details
### Design

The design of the TFTC Calibration platform is shown in Figure 1. A serpentine-shaped Pt microheater is designed at the bottom of the substrate and it forms three distinct heating zones in which TFTC junction is placed. Pt-based microheaters are commonly used as heating elements for on-chip heating purposes [25-27]. The purpose of having a serpentine-shaped heater is to allow the heat to uniformly distribute in the region and provide a hot zone where the TFTC junction can be placed for calibration. Platinum and chromium were chosen to be the two metals of the TFTC because of their large difference in Seebeck coefficient, and compatibility with our fabrication process. Before fabricating micro-TFTCs, mechanical and thermoelectric simulations were carried out to see the thermal behavior of the calibration platform and choose the appropriate substrate for the experiment.

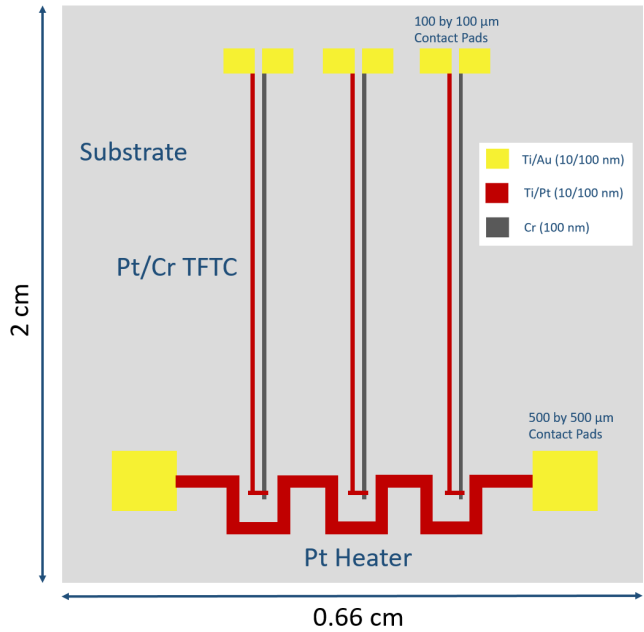

Figure 1 Top view of Pt-Cr thin-film thermocouple structure and calibration platform used for Simulation and Experiments

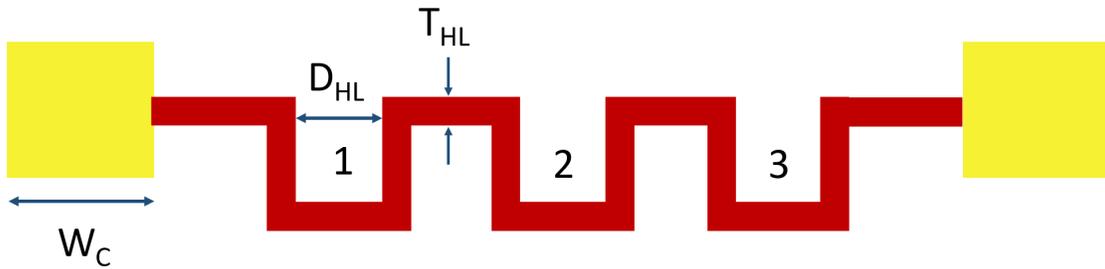

Figure 2 Microheater design with dimensions, Heating zones marked as 1, 2 , 3 are the locations for the placement of TFTC junction

Figure 2 gives the details of each dimension and values are presented in Table 1. Three distinct (marked as 1, 2, 3) heating zones are used to form hot junctions.

Table 1 Dimensions of Pt-based Microheater

| Notation | Description | Dimensions |
|----------|-------------|------------|
|          |             |            |

| $T_{HL}$ | Thickness of Heating Line | 100 µm |
| --- | --- | --- |
| $W_C$ | Width of Contact pad (Square Pad) | 500 µm |
| $D_{HL}$ | Distance between two legs of the heater | 250 µm |

Similarly, the dimensions of TFTC are shown in Figure 3 and values are presented in Table 2. The junction is clearly shown by dashed lines where two metals overlap each other. Thermocouples with varying layouts and sizes ranging from 1 µm to 10 µm were fabricated and tested to determine their performance characteristics.

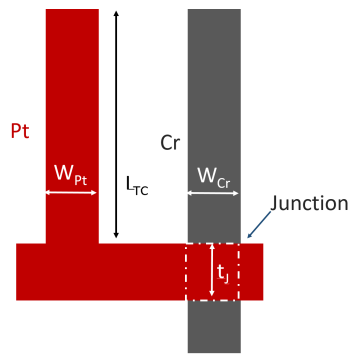

Figure 3 Thin film Thermocouple Design with dimensions

Table 2 Description of TFTC Design Parameters

| Notation | Description | Dimensions |
| --- | --- | --- |
| $t_J$ | TFTC Junction Width | 1 µm, 2 µm, 5 µm, 7 µm, 10 µm |
| $L_{TC}$ | Length of TFTC | 16.325 mm |

**Simulations**

Numerical simulations were carried out using COMSOL Multiphysics by designing a three-dimensional model of the calibration platform. A 3D model of the calibration platform was designed with dimensions as shown in Figure 1. In the thermoelectric simulation, the boundary conditions of the platform were set according to the real laboratory test environment. The microheater and left TFTC electrode were set as platinum, and the right TFTC electrode was set as chromium. Contact pads were set as Titanium/ Gold. Convective heat flux conditions were used for analysis with the ambient set as Air and the ambient temperature was set to 25 °C. A printed circuit board (PCB) was placed below the substrate to mimic the laboratory conditions. Silicon, Sapphire, and Borosilicate glass were used as substrate materials for the comparison of temperatures in hot and cold zones. Voltages were applied to the microheater and a corresponding rise in temperature in the heated zone was compared to the cold zone of the substrate. The substrate thickness was set at 500 microns. A standard thickness of 1.6 mm was used for PCB in the simulation. Only 3$^{rd}$ of the actual chip dimensions was simulated due to symmetry.

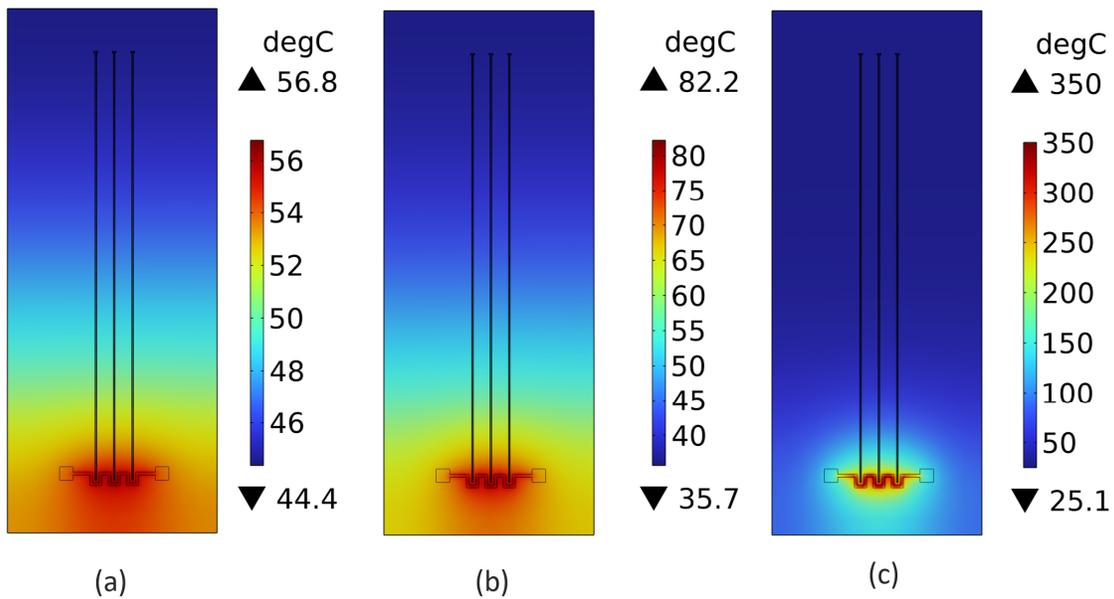

Figure 4 Temperature Distribution in different substrates

Figure 4 shows temperature distribution across Silicon, Sapphire, and Borosilicate glass substrates respectively when a voltage of 10 V is applied to the microheater. Among these materials, Borosilicate glass exhibits the highest temperature due to its relatively low thermal conductivity (~1.38 W/mK). This characteristic of Borosilicate glass impedes the efficient transfer of heat across its surface, resulting in a higher accumulation of temperature compared to Silicon and Sapphire thus forming a heating zone with high temperature. Silicon and Sapphire, on the other hand, due to their higher thermal conductivity demonstrate a lower rise in temperature under the same conditions. Thus, the temperature distribution is more uniform in this case as shown in Figure 4 part (a) and (b). This comparison highlights the critical role of thermal conductivity in determining the temperature behavior of substrates, revealing Borosilicate glass as the best candidate to provide local heating effect. Furthermore, the temperature in the junction area ($T_J$) was analyzed. The junction is defined as the overlapping area of Pt and Cr TFTC electrodes and is shown in Figure 3 by the dashed area. Table 3 highlights the average temperatures in both the junction ($T_J$) and the cold zone ($T_C$) for all three substrates. Silicon exhibits a junction temperature of 55.805°C and a cold zone temperature of 44.464 °C, while Sapphire has a junction temperature of 77.621°C and a cold zone temperature of 35.937°C. Notably, Borosilicate Glass stands out with a significantly higher junction temperature of 269.90°C, yet its cold zone temperature is the lowest at 25.073°C. This analysis indicates that among the three substrates, only Borosilicate glass's cold zone temperature matches with room temperature conditions, which are generally considered to be in the range of 20-25°C. In contrast, the cold zone temperatures for both Silicon and Sapphire are elevated above this range as indicated before. Therefore, Borosilicate Glass demonstrates that it is an excellent choice for on-chip calibration of TFTCs.

Table 3 Temperatures in Hot and Cold zones for all three substrates ($t_J$=1 µm)

| Temperature (°C) | Silicon | Sapphire | Borosilicate Glass |
|---|---|---|---|
| $T_J$ (Average Temperature in the junction) | 55.805 | 77.621 | 269.90 |
| $T_C$ (Average temperature in the cold zone) | 44.464 | 35.937 | 25.073 |

**Device Fabrication**

Testing devices with (Pt/Cr) micro-TFTCs were fabricated on 2 cm by 2 cm borosilicate glass substrates (Borofloat 33 provided by University Wafer, Inc.) with standard cleanroom techniques for semiconductor microelectronics. This type of substrate is used for various applications such as optoelectronic devices, photonics, and microelectronics, including photovoltaics, LED devices, and thin-film solar cells. Glass substrate was first cleaned and degreased with acetone followed by isopropanol and subsequently rinsing in DI water followed by Piranha cleaning to remove any organic contaminants. Lithography of the micro-TFTC patterns was completed by maskless lithography using DWL 66$^+$ laser lithography tool, which enables high-resolution direct pattern writing. Pt and Cr (each with a thickness of 100 nm) thin films were deposited using a magnetron sputtering system. Residual photoresist was removed using oxygen plasma at 250 W for 30 seconds before film deposition. Titanium adhesion layer (of thickness 10 nm) was deposited before Pt deposition to improve adhesion. To further reduce electrical noises any leakage, a 10-nm thick $Al_2O_3$ insulating layer was deposited on top of the micro-TFTC junction region with the Atomic layer deposition. Micro-TFTCs were calibrated by fabricating a serpentine-shaped Platinum heater having a thickness of 100 nm. As in the case of TFTC, a titanium layer of 10 nm thickness was deposited to improve adhesion. Figure 5 (a) shows the microscopic image of the microheater with TFTCs placed inside the heating zones. The serpentine shape allows the heat to uniformly distribute in the region and it provides a hot zone where TFTC junction can be placed for calibration. Figure 5 (b) shows the zoomed-in image of an individual heating zone. The junction width, $t_J$ for the calibration samples was varied from 1 µm to 10 µm, while the TFTC length, $L_{TC}$, was kept at 16.325 mm. In addition, dummy thermocouples were fabricated and tested to evaluate any potential effects. During calibration, the cold ends of the samples were kept at a stable temperature of 25 °C. The heater was calibrated first. Direct Current (DC) was passed through the heater. Since the electrical power is dissipated in the heater due to Joule heating, the temperature of the substrate rises and heating zones are formed. The temperature of the heating zones was increased from 25 °C to 450 °C. Infrared Thermal Graphs of the microheater were obtained using the Optotherm Thermal Imaging System for Microelectronics Thermal Analysis. The information obtained by Infrared thermal graphs helps to find the absolute temperature in the heating zone and location of the TFTC junction. TFTC measurements were done with a Keithley 2182A nanovoltmeter and a computerized data acquisition system using LabVIEW was used to collect the Seebeck Voltage data.

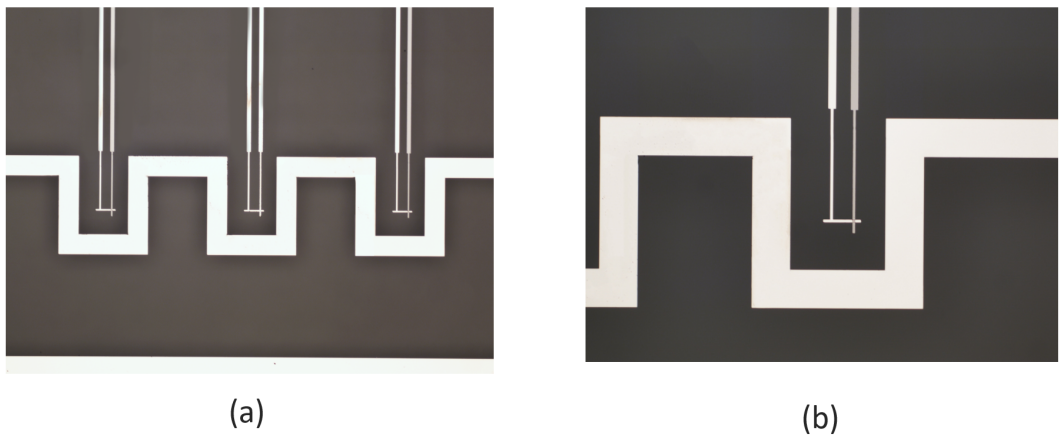

Figure 5 Device Micrograph (a) Fabricated TFTC in the heating zones (b) Zoomed-in image of a Heating zone, TFTC junction can be seen

## Results and Discussion

The first step was to calibrate the microheater. Gold wire bonds were used to connect the contact pads with a PCB and voltage was applied using a DC voltage source. As the voltage was increased, the corresponding heater temperature was measured and recorded. Figure 6 shows the IR micrograph of temperature taken by the Optotherm Thermal Imaging System. Voltage was swept from 0 to 20 V and the corresponding temperature in each heating zone was measured and is shown in Figure 7. Optotherm Thermal Imaging System allows to determine of temperatures in the view domain at each pixel using a line or box. To find the temperature at the location where TFTC would be fabricated, (a measurement box as shown in Figure 7 inset) was drawn. Temperature was averaged in that location and used as the Junction temperature, $T_J$ for subsequent calibration of TFTC.

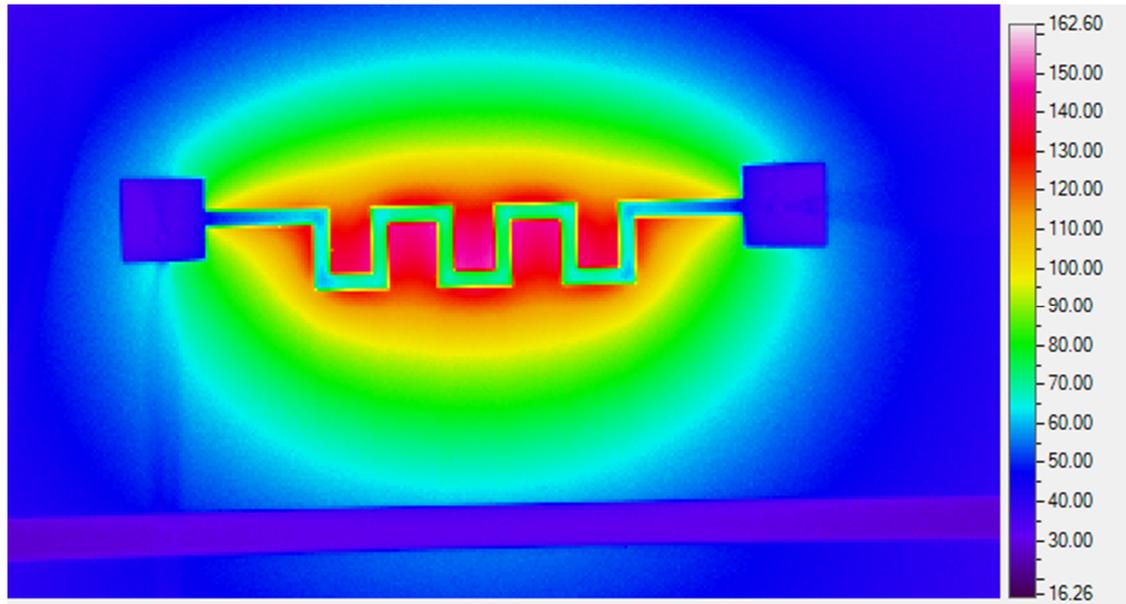

Figure 6 Optotherm Micrograph for microheater temperature

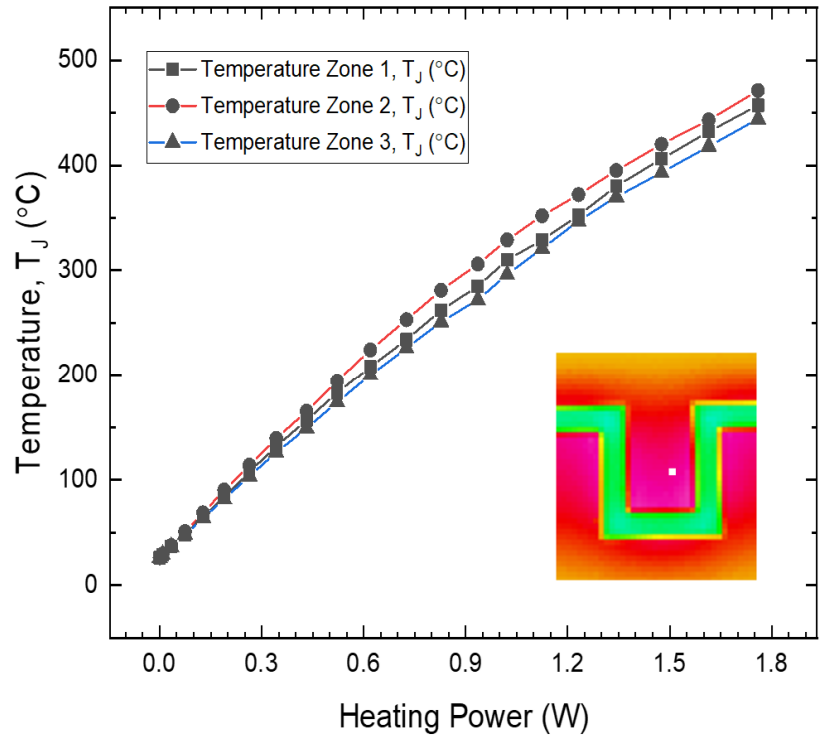

Figure 7 Junction Temperatures in heating zones with respect to heating power, the inset image shows the location of the temperature averaging spot of heating zone 2

TFTCs with different junction thicknesses were first calibrated to study how the junction sizes influence their thermal sensitivities. The spatial resolution of TFTC is determined by its junction size. Figure 8 plots the calibration results of the Pt/Cr TFTCs. The calibration results for a bunch of Pt/Cr TFTCs, with a thickness of 100 nm for Pt and Cr stripes, are presented with junction widths of 1 μm, 2 μm, 5 μm, 7 μm, and 10 μm in Figure 8 (a),(b),(c),(d) and (e) respectively. The results indicate that the TFTC has a linear relationship between temperature and Seebeck voltage up to 200 °C(R-squared= 0.997 for all TFTCs). Beyond 200 °C, the Seebeck Voltage is not linear anymore. It is due to a change in Seebeck Coefficient at high temperatures which is a common phenomenon in TFTCs [28-31]. At higher temperatures, electron scattering becomes more significant. This increased scattering affects the energy levels of charge carriers (electrons or holes). Consequently, the Seebeck coefficient changes due to alterations in charge carrier behavior. The calculated Seebeck Coefficient of Pt/Cr TFTCs is $19.23 \pm 0.405$ µV/C. It is an average value of 10 different TFTCs with a standard deviation of 2.06%. As mentioned before, dummy thermocouples were fabricated alongside the TFTC devices to ensure the accuracy and reliability of the findings. Heating zone 2 was used for dummy TFTCs. These dummy devices were fabricated with the junction based on the same metal (Pt or Cr), unlike the standard thermocouples that require two different metals to generate a voltage through the Seebeck effect. The purpose of introducing these dummy thermocouples was to confirm that any observed voltage in the experiments resulted truly from the thermoelectric effect and not from any leakage phenomenon. As expected, no voltage generation was observed in the dummy thermocouples as shown in Figure 8 which aligns with the theoretical understanding that the Seebeck effect cannot occur without a temperature difference across two dissimilar materials. This highlights the effectiveness of our calibration platform.

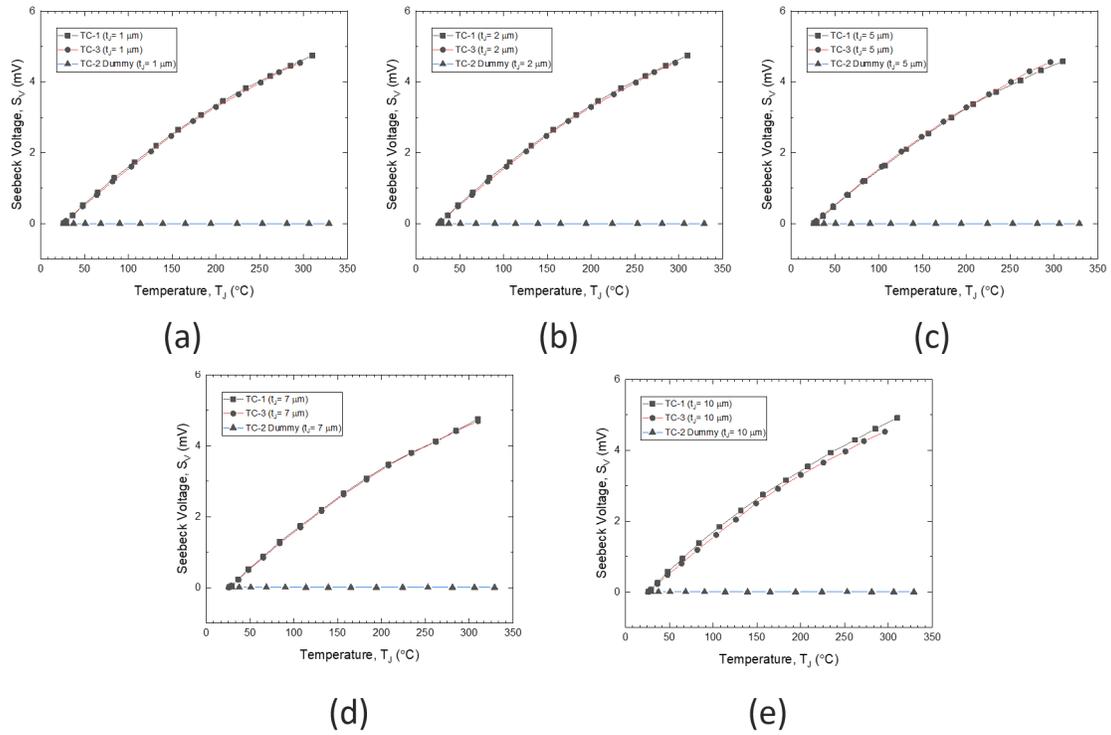

Figure 8 Seebeck Voltage measured with junction temperature $T_J$ for various TFTC junction widths, $t_J$, along with dummy TFTC response (a) 1 μm (b) 2 μm (c) 5 μm (d) 7 μm (e) 10 μm (f) Junction temperatures in heating zones

Seebeck Voltage produced by the thermocouples was found to be independent of junction size, as observed through the testing procedure. Figure 9 shows that the Seebeck Voltage of these TFTCs is insensitive to the junction width. This results is in agreement with previously published works [6, 32].

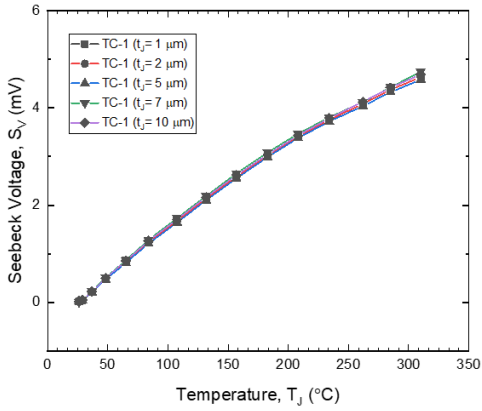 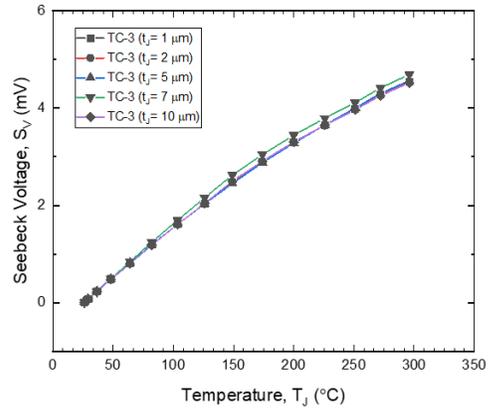

(a)            (b)

Figure 9 Seebeck Voltage vs Temperature for (a) TFTCs in heating zone 1 (b) TFTCs in heating zone 2

Our work has shown that by controlling the ambient and junction temperature of the characterization platform, TFTC with different sizes can be calibrated and the relative Seebeck coefficient can be measured with high accuracy.

## Conclusion

In this work, we precisely calibrated and measured the response of Pt/Cr-based micro-thin film thermocouples (TFTCs). We designed an improved on-chip characterization platform that includes microheater and micro-TFTCs fabricated on a borosilicate glass substrate. Pt-based microheaters fabricated on the substrate were used to calibrate micro-TFTCs. Due to the low thermal conductivity of the substrate, an elevated temperature at the TFTC junction was realized while maintaining ambient temperature (cold zone) at the open ends. The layout of the characterization platform was designed to ensure an accurate temperature measurement of the hot junction of the thermocouple and to make sure that the cold zone stays at room temperature. Numerical simulations were used to engineer both the microheater and TFTC junction for precise temperature control. TFTCs with different junction sizes (from 1 μm to 10 μm) were fabricated and characterized. The calculated sensitivity of Pt/Cr TFTCs using this calibration method is 19.23 ± 0.405 µV/C. The findings from this work demonstrate the feasibility of fabricating and calibrating high-performance micro-thin film thermocouples on a borosilicate glass substrate with high accuracy and lay the groundwork for future research into the development of miniaturized thermal sensors. These sensors could revolutionize temperature monitoring in confined spaces, offering high resolution and accuracy that are critical for the advancement of microelectronic devices and systems.

## List of References